\documentstyle[12pt]{article}
\newcommand{\beq}{\begin{equation}}
\newcommand{\eeq}{\end{equation}}
\newcommand{\bea}{\begin{eqnarray}}
\newcommand{\eea}{\end{eqnarray}}
\newcommand{\un}{\underline}
\newcommand{\half}{{\scriptstyle{{1\over 2}}}}
\newcommand{\twth}{{\scriptstyle{{1\over 12}}}}
\newcommand{\thalf}{{\scriptstyle{{3\over 2}}}}
\newcommand{\tfour}{{\scriptstyle{{3\over 4}}}}

\newcommand{\third}{{\scriptstyle{{1\over 3}}}}

\newcommand{\real}{\relax{\rm I\kern-.18em R}}

\newcommand{\quat}{{\rm I \! H}}

\newcommand{\id}{\mbox{$id$}}

\newcommand{\Tr}{\mbox{\,Tr\,}}
\newcommand{\ad}{{\rm ad}}
\newcommand{\diag}{{\rm diag}}

\newcommand{\norm}[1]{\left\| #1 \right\|}
\newcommand{\sgbar}{\bar{\sg}}
\newcommand{\tr}{\mbox{\,tr\,}}
\newcommand{\al}{\alpha}
\newcommand{\Gm}{\Gamma}

\newcommand{\dl}{\delta}

\newcommand{\veps}{\varepsilon}

\newcommand{\lm}{\lambda}
\newcommand{\Lm}{\Lambda}
\newcommand{\sg}{\sigma}

\newcommand{\Om}{\Omega}
\newcommand{\Ss}[1]{\mbox{$\cal #1$}}
\newcommand{\pr}{\partial}
\newcommand{\ddt}{\frac{d}{dt}}
\newcommand{\Lkw}{\vec{L}_1^2}
\newcommand{\ip}[1]{\left\langle #1 \right \rangle }
\newcommand{\Order}[1]{\Ss{O}\left(#1\right)}

\begin{document}
\vskip-1cm
\hfill INLO-PUB-12/93
\vskip5mm
\begin{center}
{\LARGE{\bf{\underline{Zooming-in on the SU(2) fundamental}}}}\\
{\LARGE{\bf{\underline{domain}}}}\\
\vspace*{1cm}{\large Pierre van Baal and Bas van den Heuvel} \\
\vspace*{1cm}
Instituut-Lorentz for Theoretical Physics,\\
University of Leiden, PO Box 9506,\\
NL-2300 RA Leiden, The Netherlands.\\ 
\end{center}
\vspace*{5mm}{\narrower\narrower{\noindent
\underline{Abstract:}  For SU(2) gauge theories on the three-sphere
we analyse the Gribov horizon and the boundary of the fundamental 
domain in the 18 dimensional subspace that contains the tunnelling 
path and the sphaleron and on which the energy functional is 
degenerate to second order in the fields. We prove that parts of 
this boundary coincide with the Gribov horizon with the help of
bounds on the fundamental modular domain.}\par}

\section{Introduction} From a perturbative point, 
the Hamiltonian formulation of
gauge theories is cumbersome, and the covariant path integral
approach of Feynman is vastly superior. This remains true for
certain non-perturbative features, like instanton contributions,
which vanish to all orders in perturbation theory, and are
determined by expanding around euclidean (classically forbidden)
solutions by means of semiclassical or steepest descent
approximations. When non-perturbative effects will be important
and start to affect quantities that do not vanish
perturbatively, the method breaks down dramatically~\cite{baa1,baa2}.  
When this happens, the Hamiltonian formulation becomes superior,
especially as long as only for a limited set of low-lying energy
modes non-perturbative effects become appreciable. This has been
our strategy in dealing with gauge theories in a finite volume.
Due to asymptotic freedom, keeping the volume small allows us to
keep the number of modes which behave non-perturbatively low. An
essential feature of the non-perturbative behaviour is that the
wave functional spreads out in configuration space to become
sensitive to its non-trivial geometry. If wave functionals are
localized within regions much smaller than the inverse curvature
of the field space, the curvature has no effect on the wave
functionals.  At the other extreme, if the configuration space has
non-contractable circles, the wave functionals are drastically
affected by the geometry, or topology, when the support extends
over the entire circle (i.e.  bites in its own tail). we know
from Singer~\cite{sin} that the topology of the Yang-Mills configuration
space $\cal{A}/\cal{G}$ ($\cal{A}$ is the collection of connections,
$\cal{G}$ the group of local gauge transformations) is highly
non-trivial.  It also has a Riemannian geometry~\cite{bab} that can be
made explicit, once explicit coordinates are chosen on
$\cal{A}/\cal{G}$.

The geometry of the finite volume, which will be considered in
this paper, is the one of a three-sphere~\cite{cut1}. The general arguments
are of course independent of this geometry, in which case we will
denote (compactified) three-space by $M$. Nevertheless, the
details of the way $\cal{A}/\cal{G}$ is parametrized will
crucially depend on $M$. This is already evident from Singer's
argument~\cite{sin} as the topology of $\cal{A}/\cal{G}$ does depend on
$M$. We will come back to the consequences of this for the
physics of the problem at the end of this paper. The physical
interpretation of a Hamiltonian~\cite{chr} is clearest in the Coulomb 
gauge, $\pr_i A_i=0$. But it has been known since Gribov's work~\cite{gri}
that this does not uniquely fix the gauge. Furthermore, there are
coordinate "singularities" where the Faddeev-Popov determinant
vanishes. Here the mapping between $\cal{A}/\cal{G}$ and the
transverse vector potentials becomes degenerate.

\section{Gribov and fundamental regions}
Like using stereographic coordinates for a sphere, which leads to
a coordinate singularity at one of the poles, coordinate
singularities can be removed at the price of having different
coordinate patches with transition functions at the overlaps. In
gauge theory, these different coordinate patches can simply be
seen as different gauge choices~\cite{baa2,baa5}. But this is somewhat 
cumbersome to formulate and most, but (as we shall see) not all coordinate
singularities can be avoided if one restricts the set of
transverse vector potentials to a fundamental region which
constitutes a one to one mapping with $\Ss{A}/\Ss{G}$. This is achieved by
minimizing the $L^2$ norm of the vector potential along the gauge
orbit~\cite{sem,del1}
\beq
\norm{^g A}^2~\equiv~ -\int_M d^3x~
\tr \left( \left( g^{-1} A_i g + g^{-1} \pr_i g \right)^2\right),
\label{gAnorm}
\end{equation}
where the vector potential is taken anti-hermitian. For $SU(2)$,
in terms of the Pauli matrices $\tau_a$, one has:
\bea
A_i (x)&=& iA^a_i (x) \frac{\tau_a}{2}~,\nonumber\\
g (x) &=& \exp\left(X (x)\right),~X(x) = iX^a (x)
\frac{\tau_a}{2}.
\eea
Expanding around the minimum of eq.~(\ref{gAnorm}), one easily finds:
\bea
\norm{^g A}^2 &=& \norm{A}^2+2\int_M \tr(X
\partial_i A_i)+\int_M \tr (X^\dagger FP (A) X) \nonumber \\
&&+\frac{1}{3}\int_M\tr\left(X\left[[A_i,X],\partial_i X\right]\right)
+\frac{1}{12}\int_M\tr\left([D_iX,X][\partial_i X,X]\right)+\Order{X^5}.
\label{Xexpansie}
\eea
Where $FP(A)$ is the Faddeev-Popov operator $(\ad(A)X~\equiv~[A,X])$
\beq
FP (A)~=~-\partial_i D_i (A)~=~-\partial^2_i -\partial_i\ad(A_i).
\label{FPdef}
\eeq
At the absolute minimum the vector potential is hence transverse,
$\partial_i A_i~=~0$, and $FP(A)$ is a positive operator. The
set of all transverse vector potentials with positive Faddeev-
Popov operator is by definition the Gribov region $\Omega$. It is
a convex subspace of the set of transverse connections $\Gamma$,
with a boundary $\partial \Omega$ that is called the Gribov
horizon. At the Gribov horizon, the \underline{lowest} eigenvalue
of the Faddeev-Popov operator vanishes, and points on
$\partial\Omega$ are hence associated with coordinate
singularities. Any point on $\partial\Omega$ has a finite
distance to the origin of field space and in some cases even
uniform bounds can be derived~\cite{del2,zwa1}. 

The Gribov region is the set of
\underline{local} minima of the norm functional (3) and needs to
be further restricted to the absolute minima to form a
fundamental domain, which will be denoted by $\Lambda$. The
fundamental domain is clearly contained within the Gribov region and can
easily be shown to also be convex~\cite{sem,del1}. 
Its interior is devoid of gauge
copies, whereas its boundary $\partial\Lambda$ will in general
contain gauge copies, which are associated to those vector
potentials where the absolute minima of the norm functional are
degenerate~\cite{baa3}. If this degeneracy is continuous one necessarily has
at least one zero eigenvalue for $FP(A)$ and the Gribov horizon
will touch the boundary of the fundamental domain at these
so-called singular boundary points. By singular we mean here a
coordinate singularity. There are so-called reducible
connections~\cite{don}, and $A=0$ is the most important example, which are
left invariant by a subgroup of $\cal{G}$. As here $\cal{G}$
does not act transitively, $\cal{A}/\cal{G}$ has curvature
singularities at these reducible connections. They can be 
``blown-up'' by not dividing by their stabilizer.
For $S^3$ one can proof $A=0$ is the only such a
reducible connection in $\Lm$. (Note $\cal{G}$ is the set of all
gauge transformations, \underline{including} those that are
homotopically non-trivial). The stabilizer of $A=0$ is the group
$G(=SU(2))$ of constant gauge transformation. This gauge degree
of freedom is \underline{not} fixed by the Coulomb gauge
condition and therefore one still needs to divide by $G$ to get
the proper identification
\beq
\Lm/G~=~ \cal{A}/\cal{G}
\eeq

Here $\Lm$ is considered to be the set of absolute minima modulo
the boundary identifications, where the absolute minimum might be
degenerate. It is these boundary identifications that restore
the non-trivial topology of $\cal{A}/\cal{G}$. Furthermore, the
existence of non-contractable spheres allows one to prove that
singular boundary points cannot be avoided~\cite{baa3}. However, not all
singular boundary points, even those associated with continuous
degeneracies, need to be associated with non-contractable
spheres. Note that absolute minima of the norm functional are
degenerate along the constant gauge transformations, this is a
trivial degeneracy, also giving rise to trivial zero-modes for
the Faddeev-Popov operator, which we ignore. The action of $G$ is
essential to remove the curvature singularities mentioned above
and also greatly facilitates the standard Hamiltonian formulation
of the theory~\cite{chr}. There is no problem in dividing out $G$ by
demanding wave functionals to be gauge singlets (colourless states)
with respect to $G$. In practice this means effectively that one
minimizes the norm functional over $\Ss{G}/G$.

\section{Gauge fields on the three-sphere}
We will now specialize to the case of $S^3$, for which we will summarize
the formalism that was developed in~\cite{baa1}.
We embed $S^3$ in $\real^4$ by considering the unit sphere parametrized by a 
unit vector $n_\mu$. We introduce the unit quaternions $\sg_\mu$ and their
conjugates $\bar{\sg}_\mu = \sg^\dagger_\mu$ by
\beq
  \sg_\mu = ( \id ,i \vec{\tau}), \hspace{1.5cm} 
  \bar{\sg}_\mu = ( \id,- i \vec{\tau}).
\eeq
They satisfy the multiplication rules
\beq
 \sg_\mu \sgbar_\nu = \eta^\al_{\mu \nu} \sg_\al, \hspace{1.5cm}
  \sgbar_\mu \sg_\nu = \bar{\eta}^\al_{\mu \nu} \sg_\al,
\eeq
where we used the 't Hooft $\eta$ symbols~\cite{tho}, generalised slightly to
include a component symmetric in $\mu$ and $\nu$ for $\al=0$.
We can use $\eta$ and $\bar{\eta}$ to define orthonormal framings of $S^3$,
which were motivated by the particularly simple form of the instanton 
vector potentials in these framings. The framing for $S^3$ is 
obtained from the framing of $\real^4$ by restricting in the following 
equation the four-index $\al$ to a three-index $a$
(for $\al = 0$ one obtains the normal on $S^3$):
\beq
  e^\al_\mu = \eta^\al_{\mu \nu} n_\nu , \hspace{1.5cm}
  \bar{e}^\al_\mu = \bar{\eta}^\al_{\mu \nu} n_\nu.
\eeq

The orthogonal matrix $V$ that relates these two frames is given by
\beq
  V^i_j = \bar{e}^i_\mu e^j_\mu = \frac{1}{2} 
  \tr ( (n \cdot \sg) \sg_i (n \cdot \bar{\sg})\sg_j ).
\eeq
Note that $e$ and $\bar{e}$ have opposite orientations.
Each framing defines a differential operator
\beq
  \pr^i = e^i_\mu \frac{\pr}{\pr x^\mu} , \hspace{1.5cm}
  \bar{\pr}^i = \bar{e}^i_\mu \frac{\pr}{\pr x^\mu},
\eeq
to which belong $SU(2)$ angular momentum operators, which for
historical reasons will be denoted by $\vec{L}_1$ and $\vec{L}_2$:
\beq
  L_1^i = \frac{i}{2}~\pr^i , \hspace{1.5cm}
  L_2^i = \frac{i}{2}~\bar{\pr}^i.
\eeq
They are easily seen to satisfy the condition 
\beq
  \Lkw = \vec{L}_2^2.
\eeq

The (anti-)instantons~\cite{bel} in these framings, obtained from those
on $\real^4$ by interpreting the radius in $\real^4$ as the
exponential of the time $t$ in the geometry $S^3 \times \real$, become
($\vec\veps$ and $\vec A$ are defined with respect to the 
framing $e^a_\mu$ for instantons and with respect to the framing
$\bar{e}^a_\mu$ for anti-instantons)
\beq
  A_0 = \frac{\vec{\veps} \cdot \vec{\sg}}{2 ( 1 + \veps \cdot n )}
  , \hspace{1.5cm}
  A_a = \frac{\vec{\sg} \wedge \vec{\veps} -( u + \veps \cdot n ) 
  \vec{\sg}} {2 ( 1 + \veps \cdot n )},\label{vecA}
\eeq
where 
\beq
   u = \frac{ 2 s^2}{1 + b^2 + s^2} , \hspace{1.5cm}
   \veps _\mu = \frac{2 s b_\mu}{1 + b^2 + s^2} , \hspace{1.5cm}
   s = \lm e^t.
\eeq
The instanton describes tunnelling from $A = 0$ at $t = - \infty$ to
$A_a = - \sg_a$ at $t = \infty$, over a potential barrier that is
lowest when $b_\mu \equiv 0$. This configuration (with $b_\mu = 0$, $u = 1$)
corresponds to a sphaleron~\cite{kli}, i.e.\ the vector potential $A_a = 
\frac{- \sg_a}{2}$
is a saddle point of the energy functional with one unstable mode, 
corresponding to the direction ($u$) of tunnelling. At $t = \infty$, 
$A_a = - \sg_a$ has zero energy and is a gauge copy of $A_a = 0$ by a
gauge transformation $g = n \cdot \sgbar$ with winding number one, since
\beq
  n \cdot \sg \pr_a n \cdot \sgbar = - \sg_a.
\eeq

We will be concentrating our attention to the modes that are degenerate
in energy to lowest order with the modes that describe tunnelling
through the sphaleron and "anti-sphaleron". The latter corresponds
to the configuration with the minimal barrier height separating $A = 0$
from its gauge copy by a gauge transformation $g = n \cdot \sg$
with winding number $-1$. The anti-sphaleron is actually a copy of the
sphaleron under this gauge transformation, as can be seen from 
eq.~(\ref{vecA}), since
\beq
  n \cdot \sgbar e^a_\mu \sg_a n\cdot\sg = - \bar{e}^a_\mu \sg_a,
\eeq
(with which we correct a typo in eq.~(12) of~\cite{baa1}. This also
affected the sign of eq.~(83) of this reference. We stick to the 
present more natural conventions.)
The two dimensional space containing the tunnelling paths through the
sphalerons is consequently parametrized by $u$ and $v$ through
\bea
  A_\mu(u,v)&=&\left(-u e^a_\mu-v\bar{e}^a_\mu \right)\frac{\sg_a}{2}
  =A_i(u,v)e^i_\mu,\nonumber\\
  A_i(u,v)&=&\left(-u\dl^a_i-v V^a_i\right)\frac{\sg_a}{2}=
  -u\frac{\sg_i}{2}+v~n\cdot\bar{\sg}\frac{\sg_i}{2}n\cdot\sg.
\eea
The gauge transformation with winding number $-1$ is easily seen to
map $(u,v) = (w,0)$ into $(u,v) = (0,2-w)$. In particular,
as discussed above, it maps the sphaleron $(1,0)$ to the
anti-sphaleron $(0,1)$. 

The Gribov and fundamental regions will be discussed in the next section. 
After that we will investigate the 18 dimensional space defined by
\bea
  A_\mu(c,d)&=&\left(c^a_i  e^i_\mu+d^a_j\bar{e}^j_\mu \right) 
  \frac{\sg_a}{2}=A_i(c,d)e_\mu^i,\nonumber \\
  A_i(c,d)&=&\left(c^a_i+d^a_j V^j_i \right)\frac{\sg_a}{2}.\label{Acddef}
\eea
One easily verifies that the $c$ and $d$ modes
are mutually orthogonal and that $A(c,d)$ satisfies the Coulomb gauge 
condition:
\beq
  \pr_i A_i(c,d) = 0.
\eeq
This space contains the $(u,v)$ plane through $c^a_i = -u \dl^a_i$ and
$d^a_i = -v \dl^a_i$. The significance of this 18 dimensional space is that the
energy functional~\cite{baa1} 
\beq
  \Ss{V}(c,d) \equiv - \int_{S^3} \frac{1}{2} \tr(F_{ij}^2)
  = \Ss{V}(c) + \Ss{V}(d) + \frac{2 \pi^2}{3}
   \left\{ (c^a_i)^2 (d^b_j)^2 - (c^a_i d^a_j)^2 \right\}\label{pot}
,\eeq
\beq
  \Ss{V}(c) = 2 \pi^2 \left\{ 2 (c^a_i)^2 + 6 \det c + 
  \frac{1}{4}[(c^a_i c^a_i)^2 - (c^a_i c^a_j)^2 ] \right\} ,
\eeq
is degenerate to second order in $c$ and $d$. Indeed, the quadratic 
fluctuation operator \Ss{M} in the Coulomb gauge, defined by
\bea
  - \int_{S^3} \frac{1}{2} \tr(F_{ij}^2) 
  &=& \int_{S^3} \tr(A_i \Ss{M}_{ij} A_j) + \Order{A^3},\nonumber \\
  \Ss{M}_{ij} &=& 2\vec{L}_1^2\dl_{ij} + 2 \left( \vec{L}_1 + \vec{S}_{ij} 
  \right)^2,\quad S^a_{ij}=-i \veps_{aij},\label{fluct}
\eea
has $A(c,d)$ as its eigenspace for the eigenvalue $4$. It is an important 
fact~\cite{baa1}
that there is \un{one} lower eigenvalue  ($=3$) with a 12 dimensional
eigenspace.

\section{Gribov and fundamental regions for A(u,v)}
Let us analyse the condition for $\norm{^gA}^2$ to be minimal a little 
closer. We can write
\bea  
  \norm{^gA}^2 - \norm{A}^2 &=& 
     \int \tr \left( A_i^2 \right) 
     - \int \tr \left( \left( g^{-1} A_i g + g^{-1} \pr_i g \right)^2\right) 
   \nonumber \\
  &=& \int \tr \left( g^\dagger FP_\half(A)~g \right) 
       \equiv \ip{g,FP_\half(A)~g},
  \label{FPhalfdef}
\eea
where $FP_\half(A)$  is the Faddeev-Popov operator generalised to the
fundamental representation:
\beq
  FP_t(A) = - \pr_i^2 - \frac{i}{t} A^a_i T_t^a \pr_i.
  \label{FPtdef}
\eeq
Here $\vec{T}_t$ are the hermitian gauge generators in the spin-$t$ 
representation:
\beq
  T^a_\half = \frac{\tau_a}{2} , \hspace{1.5cm}
  T^a_1 = \mbox{ad}(\frac{\tau_a}{2}).
\eeq
They are angular momentum operators that
satisfy $\vec{T}^2_t = t (t+1)\id$. At the critical points
$A \in \Gm$ of the norm functional,  
(recall $\Gm = \{ A \in \Ss{A} | \pr_i A_i = 0 \}$),
$FP_t(A)$ is an hermitian operator. Furthermore, $FP_1(A)$ in that case 
coincides with the Faddeev-Popov operator $FP(A)$ in eq.~(\ref{FPdef}).

In eq.~(\ref{FPhalfdef}) $FP_\half(A)$ is defined as an hermitian
operator acting on the vector space $\Ss{L}$ of functions $g$ over $S^3$
with values in the space of the quaternions $\quat=\{q_\mu\sg_\mu|
q_\mu\in\real\}$.
To be precise, we should require $g \in W^1_2(S^3,\quat)$,
with  $W^1_2(M,V)$ the Sobolev space of functions on $M$ with
values in the vector space $V$, whose first derivative is continuous
and square integrable. We use the standard isomorphism between the 
complex spinors $\psi$ (on which $\vec T_\half$ acts in the standard
way) and the quaternions, by combining $\psi$ and $\bar{\sg}_2\psi^*$.
To be specific, if $\psi_1=q_0+iq_3$ and $\psi_2=iq_1-q_2$, then 
$g=(\psi,\bar{\sg}_2\psi^*)=q\cdot\sg$ is a quaternion 
(on which $\vec T_\half$ now acts
by matrix multiplication). Charge conjugation symmetry, ${\cal C}\psi=
\bar{\sg}_2\psi^*$, implies that $[FP_\half(A),{\cal C}]=0$ and 
guarantees that the 
operator preserves this isomorphism. Also note that this symmetry
implies that all eigenvalues are two-fold degenerate.
The gauge group $\Ss{G}$ is contained in $\Ss{L}$ by restricting
to the unit quaternions: 
$\Ss{G}=\{g\in\Ss{L}|g=g_\mu\sg_\mu,g_\mu\in\real,g_\mu g_\mu = 1 \}$.

We can define $\Lm$ in terms of the absolute minima (apart from the 
boundary identifications) over $g \in \Ss{G}$ of
$\ip{g, FP_\half(A)~g}$
\beq
  \Lm = 
  \{ A \in \Gm | \min_{g \in \Ss{G}} \ip{g, FP_\half(A)~g}=0 \}.
  \label{Lmdef}
\eeq
When minimizing the same functional over the larger space $\Ss{L}$ 
one obviously should find a smaller result, i.e.
\beq
   \Ss{G} \subset \Ss{L} \Rightarrow
   \min_{g \in \Ss{G}} \ip{g, FP_\half(A)~g} \geq 
   \min_{g \in \Ss{L}} \ip{g, FP_\half(A)~g}.
    \label{gong}
\eeq
Writing
\beq
  \tilde{\Lm} = 
  \{ A \in \Gm | \min_{g \in \Ss{L}} \ip{g, FP_\half(A)~g}=0 \},
  \label{Lmtildef}
\eeq
it follows directly from eq.~(\ref{gong}) that $\tilde{\Lm} \subset \Lm$. 
Since $\tilde{\Lm}$ is related to the minimum of a functional on
a \un{linear} space, it will be easier to analyse $\tilde{\Lm}$ than $\Lm$.
We were inspired by appendix A of ref.~\cite{zwan} for this consideration.
Remarkably, we will be able to prove that the boundary $\partial\tilde{\Lm}$
will touch the Gribov horizon $\partial\Om$. This establishes the existence
of singular points on the boundary of the fundamental domain due
to the inclusion $\tilde{\Lm} \subset \Lm \subset \Om$.

In the $(u,v)$ plane one easily finds that
\beq
  FP_t(A(u,v)) = 4 \Lkw + \frac{2}{t} u \vec{L}_1 \cdot \vec{T}_t
    + \frac{2}{t} v \vec{L}_2 \cdot \vec{T}_t.
    \label{FPuvdef}
\eeq
For fixed angular momentum $l\neq0$ (where $\Lkw = \vec{L}_2^2 = l (l+1)$),
the eigenvalues of $\vec{L}_a \cdot \vec{T}_\half$ (which is a kind
of spin-orbit coupling) are $-\frac{l+1}{2}$ and
  $\frac{l}{2}$. This is easily seen to imply that for $g$ with
$\Lkw g = l (l+1) g~~(l \neq 0)$
\beq
    -\frac{l+1}{2} \norm{g}^2 \leq \ip{g, \vec{L}_a \cdot \vec{T}_\half~g} 
    \leq \frac{l}{2} \norm{g}^2,
    \label{TLbound}
\eeq
and hence
\beq
    \ip{g, FP_\half(A(u,v))~g} \geq \norm{g}^2 
    \left[ 4 l (l+1) - (2 l + 1) ( |u| +|v| ) - u - v \right],
    \label{FPbound}
\eeq
whereas of course for $l=0$ we have $\ip{g, FP_\half(A)~g} = 0$.
Now let $\tilde{\Lm}_l$ be the region in the $(u,v)$ plane where the
right hand side of eq.~(\ref{FPbound}) is positive:
\beq
  \tilde{\Lm}_l = \left\{(u,v)| \left[ 4 l (l+1) - (2 l + 1) 
  ( |u| +|v| ) - u - v \right] \geq 0 \right\},\label{Lmtilldef}
\eeq
then one easily verifies
that $\tilde{\Lm}_l \subset \tilde{\Lm}_{l + \half}$ for $l \neq0$.
For illustration we have drawn the boudaries of  $\tilde{\Lm}_\half$ and
$\tilde{\Lm}_1$ in fig.~1 (the two nested trapeziums).
Consequently, restricted to the $(u,v)$ plane $\tilde{\Lm}_\half$,
the trapezium spanned by the four
points $(1,0)$, $(0,1)$, $(-3,0)$ and $(0,-3)$, is contained in $\tilde{\Lm}$.
As one easily checks, the vector potentials belonging to the sphalerons
at $(1,0)$ and $(0,1)$ have the same norm. Since they are related by a gauge
transformation (as was proved earlier) and lie on the boundary of 
$\tilde{\Lm}_\half$, these sphalerons have to be on the 
boundary of the fundamental domain $\partial \Lm$. Hence, $\tilde{\Lm}_\half$
is seen to provide already quite a strong bound. 

Before constructing $\tilde{\Lm}$ in the $(u,v)$ plane, it is instructive
to consider first the Gribov horizon, which is given by the zeros of
the Faddeev-Popov determinant $\det(FP_1(A))$.
The operator $FP_t(A(u,v))$ as given by eq.~(\ref{FPuvdef}) not only
commutes with $\Lkw = \vec{L}_2^2$, but also with
$\vec{J}_t$, where
\beq
  \vec{J}_t = \vec{L}_1 + \vec{L}_2 + \vec{T}_t.
  \label{Jdef}
\eeq
Using the quantum numbers $(l,j_t,j_t^z)$ one can easily diagonalise
$FP_t(A(u,v))$ for low values of $l$. Note that the eigenvalues are
independent of $j_t^z$.
Defining the scalar and pseudoscalar helicity combinations
\beq
  s= u+v , \hspace{1.5cm} p = u-v,
\eeq
we take from ref.~\cite{baa4} the results
\bea
  \det\left(FP_1(A(u,v))|_{l = \half}\right) &=& 
      \left( 3 - 2\,s \right) \,
    {{\left( 9 - 3\,s - 2\,{p^2}  \right) }^3}
    \,{{\left( 3 + s \right) }^5},
    \label{deteenhalf} \\
  \det\left(FP_1(A(u,v))|_{l = 1}\right) &=&
  512 (8 - 2s ) ( (8 -2 s )^2 - s^2  + p^2 (2 s - 7) )^3\times\nonumber \\
  && ( 64 - s^2 - 3 p^2)^5 (8 + 2s )^7,
  \label{deteeneen}
\eea
the zeros of which are also exhibited in fig.~1
(solid lines for $l=\half$, dashed lines for $l=1$).
The Gribov horizon in the $(u,v)$ plane is indicated by the fat lines and is
completely determined by the $l=\half$ sector, a fact that we will now prove.
Note that the set of infinitesimal gauge transformations
$L_g = \{ X: S^3 \rightarrow su(2) \}$, where $su(2)$ is the Lie-algebra
for $SU(2)$ (i.e. the traceless quaternions), is contained in \Ss{L}. It is 
easy to verify that for $X \in L_g$, we have for all vector potentials $A$
\beq
  \ip{X, FP_1(A)~X} = \ip{X, FP_\half(A)~X}.\label{halfeen}
\eeq
This fact will enable us to use the same bounds for $FP_1$ and $FP_\half$
(cf.\ eq.~(\ref{gong})):
\beq
   L_g \subset \Ss{L} \Rightarrow\min_{X \in L_g} \ip{X, FP_1(A)~X} \geq 
   \min_{g \in \Ss{L}} \ip{g, FP_\half(A)~g}.\label{Xong}
\eeq
Hence all zeros of the Faddeev-Popov determinant with $l\geq 1$ lie
outside the trapezium $\tilde{\Lm}_1$, spanned by the four points
$(2,0)$, $(0,2)$, $(-4,0)$, $(0,-4)$. 

This proves that $FP_1(A)\geq 0$ within the region bounded by the zeros
of eq.~(\ref{deteenhalf}). We see from fig.~1 that along 
the line $s = u+v = 3$, for $|p| = |u-v| \leq 3$, the Gribov horizon
coincides with $\partial \Lm$ and consequently these are singular boundary 
points.
Note that therefore it is necessary that the term third order in $X$ in
eq.~(\ref{Xexpansie}) has to vanish if $FP(A)~X = 0$. As on the Gribov
horizon any non-trivial zero-mode has $l=\half$, whereas $A(u,v)$ has
$l=0$ or $l=1$, this third order term vanishes along the whole
Gribov horizon in the $(u,v)$ plane (all its points are therefore
bifurcation points~\cite{baa3}). It can, however, also be shown that 
these singular boundary points are \un{not} associated with 
non-contractable spheres (see app.~A).

Next we will construct $\tilde{\Lm}$ in the $(u,v)$
plane to get an even sharper bound on $\Lm$. It is by now obvious that
this will follow from finding $\det(FP_\half(A(u,v))$ in the
$l = \half$ sector. A straightforward computation yields:
  \beq
    \det\left(FP_\half(u,v)|_{l = \half}\right) = 
    9\,{{\left( 3 + s \right) }^4}\,{{\left( 3 - 2\,s - {p^2} \right) }^2},
    \label{dethalfhalf}
  \eeq
where the multiplicity of 4 comes from the $j = 3/2$ state and the
multiplicity 2 from the two $j=\half$ states in the decomposition
$\frac{1}{2} \otimes \frac{1}{2} \otimes \frac{1}{2} 
 = \frac{1}{2} \oplus \frac{1}{2} \oplus \frac{3}{2}$.
In fig.~2 the bordering parabola going through the points
$(0,-3)$, $(1,0)$, $(\tfour,\tfour)$, $(0,1)$ and $(-3,0)$, cut off by the
line $s = -3$, forms the boundary of $\tilde{\Lm}$.
As in the case of the Gribov horizon, $\tilde{\Lm}$ is completely 
determined by the $l = \half$ sector, since also the zeros of 
$\det\left(FP_\half(A(u,v))\right)$ with $l\geq 1$ lie
outside the trapezium $\tilde{\Lm}_1$.
Notice that we have now also shown that $(u,v) = (\tfour,\tfour)$ is a
singular boundary point. 

We recall that in~\cite{baa4}, 
part of $\partial \Lm$ in the $(u,v)$ plane was constructed by expanding 
around the
sphalerons, which are known to be on $\partial \Lm$. One solves for fixed
$(u,v)$ near $(0,1)$ for the extremum of $\ip{g, FP_\half(A(u,v))~g}$
with respect to $g = n \cdot \sgbar \exp (X)$, where it can be shown that
$X = - \vec{n} \cdot \vec{\sg} f(n_0)$. This leads to a second order
differential equation, solved by
\beq
  f(x) = x \sum_{j = 1} \sum_{k = 0}^{j-1} a_{j,k}(v) u^j x^{2k},
\eeq
with
\beq
  a_{1,0}(v)=\frac{2}{2+v},~ 
  a_{2,0}=\frac{-2(v^2+6v-16)}{(2+v)^3 (10+v)},~ 
  a_{2,1}(v)=\frac{4(6+v)}{(v+2)^2 (v+10)},~\cdots
\eeq
Substituting this now back in eq.~(\ref{FPhalfdef}) and
demanding equality of norms yields $v(u)$:
\bea
 v(u) &=& 
 1-{1\over9}u^2-{2\over81}u^3-{25\over2673}u^4-{1238\over264627}u^5\nonumber\\
&&-{172442\over66950631}u^6-{687429956\over457339760361}u^7+\Order{u^8},
\eea
giving the part of $\partial \Lm$ in the $(u,v)$ plane going through
the anti-sphaleron at $u=0$. We have drawn the maximal extension to the
Gribov horizon, but not all of it is expected to coincide with
$\partial\Lm$. Interchanging the two coordinates gives
the part of $\partial \Lm$ going through the sphaleron. Both parts are 
indicated by the curves in fig.~2. They are consistent
with the inclusion $\tilde{\Lm} \subset \Lm$. In this figure also lines of 
equal potential (eq.~(\ref{pot})) are drawn.

\section{Gribov and fundamental regions for A(c,d)}
We will now generalise our discussion to the 18 dimensional field space,
parametrized by $A(c,d)$ in eq.~(\ref{Acddef}).
For this case one has
\beq
  FP_t(A(c,d)) = 4 \Lkw - \frac{2}{t} c^a_i T_t^a L_1^i  
   - \frac{2}{t} d^a_i T_t^a L_2^i.
\eeq
This still commutes with $\Lkw = \vec{L}_2^2$, but for arbitrary $(c,d)$
there are in general no other commuting operators (except for
the charge conjugation operator \Ss{C} for $t=\half$).

We first calculate the analogues of the regions $\tilde{\Lm}_l$, as 
defined in eq.~(\ref{Lmtilldef}).
We decompose
\beq
  c^a_i = \sum_{j=1}^9 c_j \left(b_j\right)^a_i, \hspace{1cm}
  d^a_i = \sum_{j=1}^9 d_j \left(b_j\right)^a_i,
\eeq
with $c_j$ and $d_j$ coefficients and the set $\{b_j\}$ a basis
of $\real^{3,3}$, consisting of orthogonal matrices ($b_j^T = b_j^{-1}$)
with unit determinant. We then have:
\bea
  FP_t(A(c,d)) &=& 4 \Lkw - \frac{2}{t} c_j T_t^a \left(b_j\right)^a_i L_1^i
  - \frac{2}{t} d_j T_t^a \left(b_j\right)^a_i L_2^i \nonumber \\
&=& 4 \Lkw  - \frac{2}{t} c_j \vec{T}_t \cdot \vec{L}_{1,j}
    - \frac{2}{t} d_j \vec{T}_t \cdot \vec{L}_{2,j},\label{rot}
\eea
with $L_{k,j}^a\equiv(b_j)_i^aL_k^i$ ($k=1,2$) proper angular momentum 
operators.
As in eq.~(\ref{FPbound}), for $g$ an eigenfunction of $\Lkw$
with eigenvalue $l(l+1)$ ($l \neq 0$), we find the bound
\beq
    \ip{g, FP_\half(A(c,d))~g} \geq \norm{g}^2 
    \left[ 4 l (l+1) - (2 l + 1) \sum_{j=1}^9 \left( |c_j| + |d_j| \right)
   + \sum_{j=1}^9 \left( c_j + d_j \right) \right].
    \label{FPcdbound}
\eeq
As before, we define $\tilde{\Lm}_l$ 
as the polyhedra where the right hand side of eq.~(\ref{FPcdbound})
is positive. They are nested polyhedra, i.e. $\tilde{\Lm}_l\subset
\tilde{\Lm}_{l + \half}$. Hence we have the 
inclusion $\tilde{\Lm}_\half \subset \tilde{\Lm} \subset \Lm \subset \Om$.
If we restrict ourselves to the two dimensional subspace where
all but one of the $c_j$ ($-u$) and all but one of the $d_j$ ($-v$) are zero,
we precisely recover the situation of the previous section.
The bounds will, however, depend on the particular choice of the $b_j$ 
matrices. The sharpest bound is obtained by forming the \un{union} 
of all $\tilde{\Lm}_l$ obtained by these various choices.

We now turn to the computation of the Faddeev-Popov determinants.
In the sector $l = \half$, which is $4 (2\,t +1 )$ dimensional, the problem
of computing $\det(FP_t(A(c,d)))$ is still manageable. A suitable basis
is given by $|s_1,s_2,s_3\rangle$, with $s_i$ the eigenvalues of the
third component of the three angular momentum operators $\vec{L}_1$, 
$\vec{L}_2$ and $\vec{T}_t$. For $t=1$ it is actually more 
convenient to consider $|s_1,s_2\rangle_a$, where $a$ is the vector 
component. Using 
\bea
  L_1^\pm |s_1,s_2\rangle_a &=& ( \half \mp s_1) |- s_1,s_2\rangle_a\ ,\quad
  L_2^\pm |s_1,s_2\rangle_a = ( \half \mp s_2) |s_1, -s_2\rangle_a\ ,\nonumber\\
  L_i^3   |s_1,s_2\rangle_a &=& s_i |s_1,s_2\rangle_a\ ,\quad
  T_1^b   |s_1,s_2\rangle_a =  i \veps_{bac} |s_1,s_2\rangle_c\ ,
\eea
where as usual $L_a^\pm = L_a^1 \pm L_a^2$,
one easily writes down the matrix for $M \equiv FP_1(A(c,d))$ in this sector
($c_\pm^a \equiv c_1^a \mp i c_2^a$ and  $d_\pm^a \equiv d_1^a \mp i d_2^a$):
\bea
  M |s_1,s_2\rangle_b &=& - i \sum_{\al = \pm} \left\{ 
       (\half - \al s_1) c_\al^a \veps_{abc} |- s_1,s_2\rangle_c
     + (\half - \al s_2) d_\al^a \veps_{abc} |s_1,- s_2\rangle_c
   \right\}\nonumber\\
&&+\left( 3 \dl_{bc} - 2 i \veps_{abc} (s_1 c_3^a + s_2 d_3^a) \right)
 |s_1,s_2\rangle_c\ .
\eea
In particular for the choice
\beq
  c^a_i = x_i \dl^a_i\ , \hspace{1.5 cm} d^a_i = y_i \dl^a_i\ ,
  \label{cddiag}
\eeq
one can, with the help of Mathematica~\cite{wol}, check that the 
following holds:
\bea
  \det\left(FP_1(A(c,d))|_{l = \half}\right) &=& 
F(x_1+y_1,x_2+y_2,x_3+y_3)F(x_1-y_1,x_2-y_2,x_3+y_3)\times\nonumber\\
&&F(x_1-y_1,x_2+y_2,x_3-y_3)F(x_1+y_1,x_2-y_2,x_3-y_3),\nonumber\\
 F(\vec{z}) &\equiv& 2 \prod_i z_i - 3 \sum_i z_i^2 + 27.
 \label{deteendiag}
\eea

To obtain the result for general $(c,d)$ we first observe that we have 
invariance under rotations generated by $\vec{L}_1$ and $\vec{L}_2$
and under constant gauge transformations generated by $\vec{T}_t$, 
implying that $\det\left(FP_t(A(c,d))|_{l = \half}\right)$ is invariant
under
\beq
  c^a_i \rightarrow \left( S c R_1 \right)^a_i, \hspace{1.5 cm}
  d^a_i \rightarrow \left( S d R_2 \right)^a_i,
  \label{invar}
\eeq
with $R_1$, $R_2$ and $S$ orthogonal matrices with unit determinant
(note that the $\vec{L}_{k,j}$, introduced in eq.~(\ref{rot}), are nothing but
the $\vec{L}_k$ generators, rotated by $R_1=R_2=b_j$).
This also allows us to understand the large amount of symmetry in 
eq.~(\ref{deteendiag}), as a permutation of the $x_i$ ($y_i$) and a
simultaneous change of the sign of two of the $x_i$ ($y_i$) is the remnant 
of this symmetry, when restricted to the diagonal configurations of 
eq.~(\ref{cddiag}).
The result for the generalization of eq.~(\ref{deteendiag}) to arbitrary 
$(c,d)$ is presented in appendix B. Here we will treat the case $d=0$.
Using eq.~(\ref{invar}) we first diagonalize $c^a_i$ and
then express $F(\vec{x})$ in terms of the complete set of 
rotational and gauge invariant parameters of $c^a_i$
\beq
  \det c = \prod_i x_i,~\Tr(c c^t) = \sum_i x_i^2,
  ~\Tr(c c^tc c^t) = \sum_i x_i^4,
\eeq
which implies $F(\vec{x}) = 2 \det c - 3 \Tr(c c^t) + 27$ and
\beq
  \det\left(FP_1(A(c,0))|_{l = \half}\right) 
= \left( 2 \det c - 3 \Tr(c c^\dagger) + 27 \right)^4.
\label{deteend0}
\eeq
This can also be easily derived by constructing the three dimensional 
invariant subspace for ($c^a_i = x_i \dl^a_i$, $d=0$), spanned by the
3 vectors $n_i \sg_i$ (no sum over $i$), with respect to which the 
matrix for $M$ takes the form
\beq
  M(n_1 \sg_1,n_2 \sg_2,n_3 \sg_3) = 
 \left( \begin{array}{ccc}
   3 & x_3 & x_2 \\
   x_3 & 3 & x_1 \\
   x_2 & x_1 & 3 \\
 \end{array} \right)
 \left( \begin{array}{c}
   n_1 \sg_1 \\
   n_2 \sg_2 \\
   n_3 \sg_3 \\
 \end{array} \right),
\eeq
whose determinant coincides with $F(\vec{x})$. It is not too difficult to 
construct the 3 other three dimensional invariant subspaces with identical 
determinants.

Two special cases in this class were first considered by 
Cutkosky~\cite{cut2}:
\bea
{\mbox{I}}&:&  c^a_i = \diag(-u+y,-u+y,-u-2y),\nonumber \\
{\mbox{II}}&:&  c^a_i = \diag(-u+x,-u-x,-u).\label{cases1}
\eea
For $F$, which determines the Faddeev-Popov determinant at $l=\half$, 
we find~\cite{cut2}
\bea
F_{\mbox{I}} &=& (u+2y+3)[(u +3)( 3 - 2u) + 2 (2u -3 )y-2 y^2],\nonumber \\
F_{\mbox{II}} &=& (u +3)^2( 3 - 2u) + 2 (u -3 ) x^2.\label{cases2} 
\eea
The associated zeros are drawn respectively in figs.~3 and 4. Note that 
the $(u,y)$ plane admits a global gauge symmetry $(u,y) \rightarrow
\frac{1}{3}(4 y - u , y + 2 u )$
generated by $S = \mbox{diag}(-1,-1,1)$, which maps the vacuum at
$(u,y) = (2,0)$ to a vacuum at $(- \frac{2}{3},\frac{4}{3})$.
To conclude that these zeros coincide with the Gribov horizon, we have
to show that the Faddeev-Popov operator for all $l\geq 1$ is positive 
within the region bounded by these zeros. 
Using eq.~(\ref{Xong}), it is sufficient to show that these
zeros lie within $\tilde{\Lm}_1$, the region obtained from the bound on 
$FP_\half(A)$ in eq.~(\ref{FPcdbound}). Clearly we should try to construct
this bound by taking for $b_j$ the diagonal orthogonal matrices $\diag(1,1,1)$,
$\diag(1,-1,-1)$, $\diag(-1,-1,1)$ and $\diag(-1,1,-1)$. It turns out to be 
sufficient to consider the union of the bounds obtained by applying 
eq.~(\ref{FPcdbound}) for the four triplets of possible choice of diagonal
$b_j$. In terms of general $x_i$, this leads to the four bounds
\bea
&&4l(l+1)-\half(2l+1)(|x_1+x_2|+|x_1+x_3|+|x_2+x_3|)-x_1-x_2-x_3\geq 0,
\nonumber\\
&&4l(l+1)-\half(2l+1)(|x_1-x_2|+|x_1-x_3|+|x_2+x_3|)-x_1+x_2+x_3\geq 0,
\nonumber\\
&&4l(l+1)-\half(2l+1)(|x_1+x_2|+|x_1-x_3|+|x_2-x_3|)+x_1+x_2-x_3\geq 0,
\nonumber\\
&&4l(l+1)-\half(2l+1)(|x_1-x_2|+|x_1+x_3|+|x_2-x_3|)+x_1-x_2+x_3\geq 0.
\label{boundd0}
\eea
The union of these polyhedra respects the gauge and rotation symmetry and
we take it as the definitions of $\tilde{\Lm}_l$ for $d=0$. They are
again nested, such that it is sufficient to show that the convex regions
bounded by the zeros of the Faddeev-Popov operator in the $l=\half$
sector are contained within $\tilde{\Lm}_1$. From figs.~3 and 4 we see that
this is indeed the case, allowing the identification of $\pr\Om$ (fat curves)
and $\pr\tilde{\Lm}$ (dashed or full lines) with the zeros of respectively
$\det(FP_1(A)|_{l=\half})$ and $\det(FP_\half(A)|_{l=\half})$.

We now turn to the calculation of 
$\det\left(FP_\half (A(c,d))|_{l = \half}\right)$ which will allow
us to construct $\tilde{\Lm}$ and to 
find possible further singular points on the boundary of the 
fundamental domain. In this case the basis $|s_1,s_2,s_3\rangle$, which
was defined earlier, is a convenient one for the $l = \half$ 
invariant subspace. Using the invariance as given by eq.~(\ref{invar}),
we can take $c^a_i$ diagonal and $d^a_i$ symmetric:
\beq
  c^a_i = 
\left( \begin{array}{ccc}
x_1 & 0 & 0 \\
0 & x_2 & 0 \\
0 & 0 & x_3 \\
\end{array} \right), \hspace{1.5 cm}
d^a_i = 
\left( \begin{array}{ccc}
y_1 & z_1 & z_2 \\
z_1 & y_2 & z_3 \\
z_2 & z_3 & y_3 \\
\end{array} \right).
\eeq
With $L_a^\pm$ as before and $T_\half^\pm = T_\half^1 \pm i T_\half^2$, we
obtain the following expression:
\bea
 &&  FP_\half(A(c,d))=3+2iz_1\left(T_\half^+L_2^+-T_\half^-L_2^-\right) 
      \nonumber \\
 && - 2 (z_2 - i z_3)\left(T_\half^3 L_2^+ + T_\half^+ L_2^3 \right)
    - 2 (z_2 + i z_3)\left(T_\half^3L_2^-+T_\half^-L_2^3\right)\nonumber \\
 && - 4 x_3 T_\half^3 L_1^3
    - (x_1 + x_2)\left(T_\half^+ L_1^- + T_\half^- L_1^+ \right) 
    - (x_1 - x_2)\left(T_\half^+ L_1^+ + T_\half^- L_1^- \right) \nonumber \\
 && - 4 y_3 T_\half^3 L_2^3
    - (y_1 + y_2) \left(T_\half^+ L_2^- + T_\half^- L_2^+ \right) 
    - (y_1 - y_2) \left(T_\half^+ L_2^+ + T_\half^- L_2^- \right).
\eea
In order to express the final result in invariants,
we introduce the matrices $X$ and $Y$ via
\beq
  X^a_b = (c c^t)^a_b,~Y^a_b = (d d^t)^a_b.
  \label{XYdef}
\eeq
Using Mathematica~\cite{wol} and expressing the result in terms of traces of 
products of $X$ and $Y$, we obtain an expression which is manifestly 
invariant:
\bea
  && \det\left(FP_\half (A(c,d))|_{l = \half}\right)=\Ss{F}^2, \\
  \label{halfcd}
  && \Ss{F}\equiv 81 - 18 \Tr (X +Y) + 24 ( \det c + \det d)
     - (\Tr(X - Y))^2 + 2 \Tr((X-Y)^2). \nonumber
\eea
With this, one easily reproduces the result of eq.~(\ref{dethalfhalf}) 
($x_i = -u$, $y_i = -v$, $s=u+v$, $p = u-v$). 
Note the overall square, which is a consequence of the two-fold
degeneration of the eigenvalues due to the fact that $FP_\half$
commutes with the charge conjugation operator \Ss{C}. Such a non-trivial
commuting operator does not exist for $FP_1$, whose determinant does 
not factorise and was hence much more difficult to calculate (see app.~B).

For $d=0$ we find
\beq
\det\left(FP_\half (A(c,0))|_{l = \half}\right) =\Ss{F}^2=
\left\{ 81 - 18 \Tr X + 24 \det c - (\Tr X)^2 + 2 \Tr(X^2) \right\}^2.
\label{dethalfd0}
\eeq
In figs.~3 and 4 we have drawn $\pr \tilde{\Lm}$ obtained from the zeros of 
eq.~(\ref{dethalfd0}) for the two cases of eq.~(\ref{cases1}):
\bea
\Ss{F}_{\mbox{I}}&=&3(3+u+2y)^2(u-1)(4y-3-u),\nonumber \\
\Ss{F}_{\mbox{II}}&=&3(u+3)(u-1)[4x^2-(3+u)^2],\label{cases3}
\eea
which indeed provides further singular boundary points 
(since $\pr \tilde{\Lm} \cap \pr \Om$ is not empty). 
Also the part of $\pr \Lm$ that contains the sphaleron
is easily derived from the fact that the gauge transformation 
with winding number $-1$, $g = n \cdot \sg$, leads to
\beq
  ^gA(x_i,0) = -(2+x_i) n \cdot \sgbar \frac{\sg_i}{2} n \cdot \sg,
\label{xcopies}
\eeq
for arbitrary diagonal configurations. Equality of norms implies the
equation $\sum x_i + 3 = 0$. This means, since $\tilde{\Lm} \subset \Lm$,
that in fig.~3 the edges of $\tilde{\Lm}$ passing through the sphalerons
coincide with $\pr \Lm$, a fact that can also be concluded from
the convexity of $\Lm$. 
Hence, in fig.~3 $\tilde{\Lm}$ coincides with $\Lm$ and the line 
$u + 2 y = -3$ consists of singular boundary points. In fig.~4 it is 
not excluded that, at the dashed lines, $\pr\tilde{\Lm}$ does \un{not}
coincide with $\pr\Lm$, as was also the case for the $(u,v)$ plane, see fig.~2.
We can settle this issue by considering the embedding of the $(u,x)$ plane
within the three dimensional space of the $x_i$. 

All surfaces to be constructed have to respect the symmetries of the 
permutations and the double sign flips of the $x_i$ coordinates. We first 
consider $\tilde{\Lm}_1$, see eq.~(\ref{boundd0}), which can be seen as a 
tetrahedron spanned by the points $(4,4,4)$, $(-4,-4,4)$, $(4,-4,-4)$ and 
$(-4,4,-4)$, enlarged by adding to each face a symmetric pyramid, whose tips 
are given by the points $(-2,-2,-2)$, $(2,2,-2)$, $(-2,2,2)$ and $(2,-2,2)$ 
(corresponding to the copies of the classical vacuum at $\vec x=0$). For 
general $l$, $\tilde{\Lm}_l$ can be constructed from this twelve faced polygon 
by scaling the corners of the tetrahedron with $l+1$ and the tips of the 
pyramids with $l$, from which their nested nature is obvious. A special case 
arises for $l=\half$, where the pyramids are of zero height, such that $\tilde
{\Lm}_\half$ is a tetrahedron. It is a remarkable fact that the fundamental 
Faddeev-Popov determinant in the sector $l=\half$ (eq.~(\ref{dethalfd0})) 
vanishes on $\pr\tilde{\Lm}_\half$. As this is enclosed by $\tilde{\Lm}_1$, 
where all eigenvalues of $FP_\half(A)$ with ${l\geq 1}$ are strictly positive, 
we conclude that $\tilde{\Lm}=\tilde{\Lm}_\half$ (the tetrahedron spanned by 
$(3,3,3)$, $(-3,-3,3)$, $(3,-3,-3)$ and $(-3,3,-3)$). 
The convex region bounded by the zeros of the adjoint determinant 
(eq.~(\ref{deteend0})) can be shown to form a surface contained in 
$\tilde{\Lm}_1$ that can be visualized by stretching a rubber sheet around this 
tetrahedron, fixed at its edges and slightly inflated. 
This surface forms the Gribov horizon $\pr\Omega$, since also all eigenvalues
of $FP_1(A)$ with ${l\geq 1}$ are strictly positive inside $\tilde{\Lm}_1$.
Because of the inclusion $\tilde{\Lm}\subset\Lm\subset\Omega$, all points on 
the edges of the tetrahedron are singular boundary points. As all the faces 
of this tetrahedron contain a sphaleron, which we have proven earlier to be on 
the boundary of the fundamental domain (the edges of the tetrahedron
are singular points on the same boundary), we conclude (using the convexity of 
$\Lm$) that $\tilde{\Lm}=\Lm$. This is consistent with eq.~(\ref{xcopies}),
where equality of norms gives the equation that describes the face
of the tetrahedron through the sphaleron at $(-1,-1,-1)$, $\sum x_i+3=0$.
The other three faces follow from flipping the sign of two of the $x_i$,
which is a symmetry. In fig.~5 we have drawn the fundamental modular domain 
for $d=0$ in $\vec x$ space and in fig.~6 we give the Gribov horizon and 
the edges of $\pr\tilde{\Lm}_1$ (dashed lines).
This completes the construction of the fundamental domain for $d=0$. 

\section{Discussion}
In this paper we have analysed in detail the boundary of the 
fundamental domain for SU(2) gauge theories on the three-sphere.
We have constructed it completely for the gauge fields with $\Lkw$=0
and have provided partial results for the 18 dimensional space of
modes that are degenerate with these in energy to second order in the fields.
Especially, the interesting point of explicitly demonstrating the 
presence of singular boundary points, i.e. points where the 
boundary of the fundamental domain coincides with the Gribov horizon,
was addressed. In ref.~\cite{baa3} existence of singular boundary points 
was proven on the basis of the presence of non-contractable 
spheres~\cite{sin} in the physical configuration space $\Ss{A}/\Ss{G}$.
This does not prove that all singular boundary points are necessarily
associated with such non-contractable spheres, which we demonstrated for
the case at hand (see app.~A). It is also important to note that
it is necessary to divide $\Ss{A}$ by the set of \un{all} gauge 
transformations, including those that are homotopically non-trivial,
to get the physical configuration space. All the non-trivial topology
is then retrieved by the identifications of points on the boundary of the 
fundamental domain. Zwanziger~\cite{zwa1} (app.~E) has constructed, for the 
case of $M=T^3$, a gauge function parametrized by a two-sphere for which the 
norm functional is degenerate, but its vector potential lies outside
the fundamental domain when also the anti-periodic gauge transformations
are considered as part of $\Ss{G}$~\cite{baa3}.

As we already mentioned in the introduction, the knowledge of the 
boundary identifications is important in the case that the wave functionals
spread out in configuration space to such an extent that they become 
sensitive to these identifications. This happens at large volumes, whereas
at very small volumes the wave functional is localized around $A=0$ and
one need not worry about these non-perturbative effects. That these
effects can be dramatic, even at relatively small volumes (above a tenth of
a fermi across), was demonstrated for the case of the torus~\cite{baa2,baa5}. 
However, for that case the structure of the fundamental domain (restricted 
to the abelian zero-energy modes) is a hypercube~\cite{baa3} and deviates 
considerably from the fundamental domain of the three-sphere. 
One can hence conclude that 
something needs to happen to the structure of the theory, to avoid that the
infinite volume limit in the infrared depends on the way this limit
is taken, e.g. by scaling different geometries, like $T^3$ or $S^3$. One
way to avoid this undesirable effect is that the vacuum is unstable against 
domain formation. We have discussed this at length elsewhere and refer
the reader to refs.~\cite{baa1,baa5,baa3} for further details. 

To conclude, let us return to the issue of the singular boundary points.
Many of the coordinate singularities due to the vanishing of the Faddeev-Popov
determinant (which plays the role of the Jacobian for the change of
variables to the gauge fixed degrees of freedom~\cite{bab} in the 
Hamiltonian formulation) are screened by the boundary of the fundamental
domain. Although the singular boundary points forms a set of zero measure
in the configuration space, they can nevertheless be important for the 
dynamics. Near these points we have to choose different coordinates and
formulate the necessary transition functions to move from one to the 
other choice. It is clear that this is difficult to formulate in all
rigour in the infinite dimensional field space. As the domain formation
is anticipated to be due to the fact that the energies of the low-lying
states flow over the sphaleron energy, we can study the dynamics of the
domain formation as long as the energies of all singular boundary points 
are well above the sphaleron energy. From figs.~2, 3 and 4 we see that
this is indeed the case in the 18 dimensional subspace we have considered.
In the future we will also investigate the 12 dimensional space of lowest
energy modes (see below eq.~(\ref{fluct})) from this perspective, 
but in the higher
energy modes the tail of the wave functional will be so small at
the singular boundary points, that we need not worry about their influence
on the spectrum. In this way we have a well defined window in which the 
non-perturbative treatment of a finite number of modes will allow us to
calculate the low-lying spectrum of the theory (see ref.~\cite{baa1} for
the set-up of this analysis).

\section*{Acknowledgements}
One of us (P.v.B.) is grateful to Dick Cutkosky for extensive correspondence
on the results quoted as ref.~\cite{cut2}, and
for the collaboration in ref.~\cite{baa4}. This work was supported in part 
by a grant from ``Stichting voor Fundamenteel Onderzoek der Materie (FOM)''.

\section*{Appendix A}
In this appendix we shall demonstrate that the singular part of the boundary
of the fundamental domain in the $(u,v)$ plane does not contain points 
associated to non-contractable spheres. Such a non-contractable sphere
implies at least a one parameter gauge function $g(t)$ along which the 
norm functional is degenerate and minimal. 
We will first show that this implies that
the fourth order term in eq.~(\ref{Xexpansie}) needs to be negative. After that
we show that this is not the case for the singular boundary points under
consideration. We write
\beq
g(t)=\exp(X(t)),\quad X(t)=tX_1+t^2X_2+t^3X_3+t^4X_4+\Order{t^5}.
\eeq
For all $t$ one should have that $\pr_i(^{g(t)}A_i)=0$. 
Using the fact that 
\beq
\pr_i\ddt(^{g(t)}A_i)= \pr_iD_i(^{g(t)}A)~(g^\dagger(t)\ddt g(t)),
\eeq
one easily concludes that
$X_1$ is a zero-mode for the Faddeev-Popov operator at $t=0$, whereas
the first order term in $t$ gives the equation 
\beq
FP(A)X_2=-\half[\pr_aX_1,[X_1,A_a]].\label{Xcorr}
\eeq
By considering the inner product of both sides of this equation with $X_1$,
we conclude that it can only have a solution provided the third
order term (for $X=X_1$) in eq.~(\ref{Xexpansie})
vanishes, as should be obviously true since we
are considering $A\in\Lm$, i.e. the norm functional is at its absolute 
minimum.
We now have sufficient information to compute $\norm{^{g(t)}A}$ to fourth
order in $t$ (the explicit forms of $X_3$ and $X_4$ drop out of the expression
for this order when we use respectively that $FP(A)~X_1=0$ and $\pr_aA_a=0$):
\beq
\norm{^{g(t)}A}^2=\norm{A}^2+t^4\left\{3\ip{X_2,FP(A)~X_2}-\twth
\ip{[D_aX_1,X_1],[\pr_aX_1,X_1]}\right\}+\Order{t^5}\label{4order}
\eeq
To obtain this result we used the Jacobi identity, partial integration,
eq.~(\ref{Xcorr}) and the assumption that $X_1$ is an eigenfunction for $\Lkw$.
Since the first term is positive definite, the norm functional can only
be degenerate if the second term is negative.

We now specialize to the case $A(u,v)$ and the singular boundary points
that occur at $(u,v)=(\tfour,\tfour)$ and $u+v=-3$ for $|u-v|\leq 3$.
The zero-modes for the Faddeev-Popov operator at these configurations
are easily seen to be given by
\beq
X_1=n_aQ^{ab}\sg_b,\quad Q^{ab}=Q^{ba},\label{XQ}
\eeq
where the trace part of the symmetric (real) tensor corresponds to 
the $j=0$ (eq.~(\ref{Jdef})) zero-mode (at $u+v=\thalf$) and the traceless
part to the $j=2$ zero-mode (at $u+v=-3$). It is now straightforward to
substitute $X_1$ (eq.~(\ref{XQ})) in eq.~(\ref{4order}). After some algebra
we find
\bea
&-&\ip{[D_aX_1,X_1],[\pr_aX_1,X_1]}/2\pi^2=2\Tr(Q^4)-2(\Tr(Q^2))^2\nonumber\\
&&\quad\quad+\third(u+v)
\left[(\Tr Q)^2\Tr(Q^2)+2\Tr Q\Tr(Q^3)-(\Tr(Q^2))^2-2\Tr(Q^4)\right]
\eea
With $Q^{ab}=\delta_{ab}$ and $u+v=\thalf$ we find for the right hand side 
3, which is positive. For a traceless $Q$ we find 
$4\Tr(Q^4)-(\Tr(Q^2))^2$ at $u+v=-3$, which is likewise strictly positive.
None of the singular boundary points can therefore be associated with a
continuous degeneracy.

\section*{Appendix B}
In this appendix we will calculate $\det\left(FP_1(A(c,d))|_{l = \half}\right)$
for general $(c,d)$, thus extending the result in eq.~(\ref{deteendiag}).
We will write the result in terms of the matrices $X$ and $Y$ defined
in eq.~(\ref{XYdef}).
It is useful to introduce the following quantities:
\bea
  D_1 &\equiv& \left(\veps_{abc} c^a_i c^b_j d^c_k \right)^2 
  = \Tr Y \left( (\Tr X)^2 - \Tr(X^2) \right) + \Tr ( X^2 Y ) 
  - 2 \Tr X \Tr (X Y),  \nonumber \\
  D_2 &\equiv& \left(\veps_{abc} c^a_i d^b_j d^c_k \right)^2 
  = \Tr X \left( (\Tr Y)^2 - \Tr(Y^2) \right) + \Tr ( X Y^2 ) 
  - 2 \Tr Y \Tr (X Y),  \nonumber \\
  D_3 &\equiv& \left( \Tr (X^2) - (\Tr X)^2 \right)
     \left( \Tr (Y^2) - (\Tr Y)^2 \right) + 2 \Tr\left( (X Y)^2 \right)
     - 2 \left( \Tr (X Y) \right)^2, \nonumber \\
  F_0 &\equiv& 27 + 2 \det c + 2 \det d - 3 \Tr X - 3 \Tr Y, \nonumber \\
  F_1 &\equiv& 36 \Tr(X Y) + 24 (\Tr X - \det c) (\Tr Y - \det d)
  + 2 D_1 + 2 D_2, \nonumber \\
  F_2 &\equiv& 6 D_3  +  8 \det c \det d
       \left( 27 - \det c - \det d + 3 \Tr X + 3 \Tr Y \right) \nonumber \\
   &&  +   4 \det c \left( 9 (\Tr (Y^2) - (\Tr Y)^2) - D_2 \right) 
       +   4 \det d \left( 9 (\Tr (X^2) - (\Tr X)^2) - D_1 \right), \nonumber \\
  F_3 &\equiv&48\left(\det c\det d\right)D_3+1296\Tr\left((X Y)^2\right)
     \nonumber \\
  &&   + 1728  \left( \det c \det d \Tr ( X Y) 
       -  \det c \Tr (X Y^2) - \det d \Tr ( X^2 Y) \right) \nonumber \\
  &&  + 96 (\det c)^2  \left( 9 \Tr(Y^2) - 6 (\Tr Y)^2 
               - 2 \det d ( \Tr X + \Tr Y ) \right) \nonumber \\
  &&  + 96 (\det d)^2  \left( 9 \Tr(X^2) - 6 (\Tr X)^2 
               - 2 \det c ( \Tr X + \Tr Y ) \right) \nonumber \\
  && +   4 \left( D_1 + 4 \det c \det d - 2 (\det d)^2 - 
        12 \det c \Tr Y \right)^2 \nonumber \\
  && +   4 \left( D_2 + 4 \det c \det d - 2 (\det c)^2 - 
        12 \det d \Tr X \right)^2 \nonumber \\
  && - 16 \left( \det c + \det d \right)^2 
       \left( (\det c)^2 - 6 \det c \det d + (\det d)^2 \right), \nonumber \\
  R &\equiv& 576 ( \Tr X + \Tr Y ) \left( \Tr ( X^2 Y^2 ) 
          - \Tr \left((X Y)^2 \right) \right) \nonumber \\
  && + 576\left( \Tr(X^2YXY)-\Tr(X^3 Y^2)+\Tr(Y^2XYX)-\Tr(Y^3 X^2) \right).
\eea
In terms of this list of invariants we have
\beq
  \det\left(FP_1(A(c,d))|_{l = \half}\right) 
    = 2 (F_0^4 + F_3) - (F_0^2 + F_1)^2 - 8 F_0 F_2 + R.
  \label{deteencd}
\eeq

The significance of this expansion becomes clear when we substitute the diagonal
choices for $c$ and $d$, eq.~(\ref{cddiag}), for which
\beq
  F_0 = E_0,~F_1 = \prod_{i=1}^3 E_i,~F_2 = \sum_{i=1}^3 E_i^2,~ 
  F_3 = \sum_{i=1}^3 E_i^4,
\eeq
with
\bea
  E_0 &=& 27 + 2 \prod x_i + 2 \prod y_i - 3 \sum x^2_i - 3 \sum y^2_i, \nonumber \\
  E_1 &=& 6 x_1 y_1 - 2 y_1 x_2 x_3 - 2 x_1 y_2 y_3, \nonumber \\
  E_2 &=& 6 x_2 y_2 - 2 y_2 x_3 x_1 - 2 x_2 y_3 y_1, \nonumber \\
  E_3 &=& 6 x_3 y_3 - 2 y_3 x_1 x_2 - 2 x_3 y_1 y_2. \nonumber \\
\eea
Most important is that $R$ vanishes identically for eq.~(\ref{cddiag}).
Hence the way we came to eq.~(\ref{deteencd}) was to first extend $E_\mu$ 
to invariant combinations, which is unique up to invariant polynomials that
vanish identically for the diagonal configuration of eq.~(\ref{cddiag}).
The space of invariant polynomials with this property is 11 dimensional and
by fitting the determinant with numerical values substituted for $c$ and $d$,
one can easily (with the help of Mathematica~\cite{wol}) solve for the 11 
coefficients. The final result of eq.~(\ref{deteencd}) has then been checked 
for a large number of random choices of $c$ and $d$.

\eject
\begin{center}
\Large{\bf Figure captions}
\end{center}
\vskip1cm
{\narrower\narrower{\noindent
Figure 1: Location, for the $(u,v)$ plane of the classical vacua (large dots), 
sphalerons (smaller dots), bounds on the Faddeev-Popov operator 
for $l=\half$ and $l=1$ (short-long dashed curves), zeros of the
adjoint determinant (solid lines for $l=\half$, dashed lines for $l=1$)
and the Gribov horizon (fat sections).}\par}
\vskip1cm
{\narrower\narrower{\noindent
Figure 2: Location of the classical vacua (large dots), sphalerons (smaller 
dots), bounds on the Faddeev-Popov operator for $l=\half$ 
(short-long dashed lines), the Gribov horizon (fat sections), zeros of the 
fundamental determinant in the sector $l=\half$ (dashed curves) and part of 
the boundary of the fundamental domain (full curves). Also indicated are the 
lines of equal potential in units of $2^n$ times the sphaleron energy.
}\par}
\vskip1cm
{\narrower\narrower{\noindent
Figure 3: Location, for the $(u,y)$ plane of the classical vacua (large dots), 
sphalerons (smaller dots), bounds on the Faddeev-Popov operator 
for $l=1$ (short-long dashed curves), boundary of the fundamental domain
(solid lines) and the Gribov horizon (fat curves), as well as the
lines of equal potential.}\par}
\vskip1cm
{\narrower\narrower{\noindent
Figure 4: Location, for the $(u,x)$ plane of the classical vacua (large dots), 
sphalerons (smaller dots), bounds on the Faddeev-Popov operator 
for $l=1$ (short-long dashed curves), boundary of the fundamental domain
(solid line), zeros of the fundamental determinant for $l=\half$ (dashed lines)
and the Gribov horizon (fat curves), as well as the
lines of equal potential.}\par}
\vskip1cm
{\narrower\narrower{\noindent
Figure 5: The fundamental modular domain for constant gauge fields on $S^3$,
with respect to the ``instanton'' framing $e_\mu^a$, in the diagonal 
representation $A_a=x_a\sg_a$ (no sum over $a$). By the dots on the
faces we indicate the spalerons, whereas the dashed lines represent the
symmetry axes of the tetrahedron.}\par}
\vskip1cm
{\narrower\narrower{\noindent
Figure 6: The Gribov horizon for constant gauge fields on $S^3$,
with respect to the ``instanton'' framing $e_\mu^a$, in the diagonal 
representation $A_a=x_a\sg_a$ (no sum over $a$). The dashed lines represent
the edges of $\tilde{\Lm}_1$, which encloses the Gribov horizon, whereas
the latter encloses the fundamental modular domain, coinciding with it at
the singular boundary points along the edges of the tetrahedron of fig.~5.
}\par}
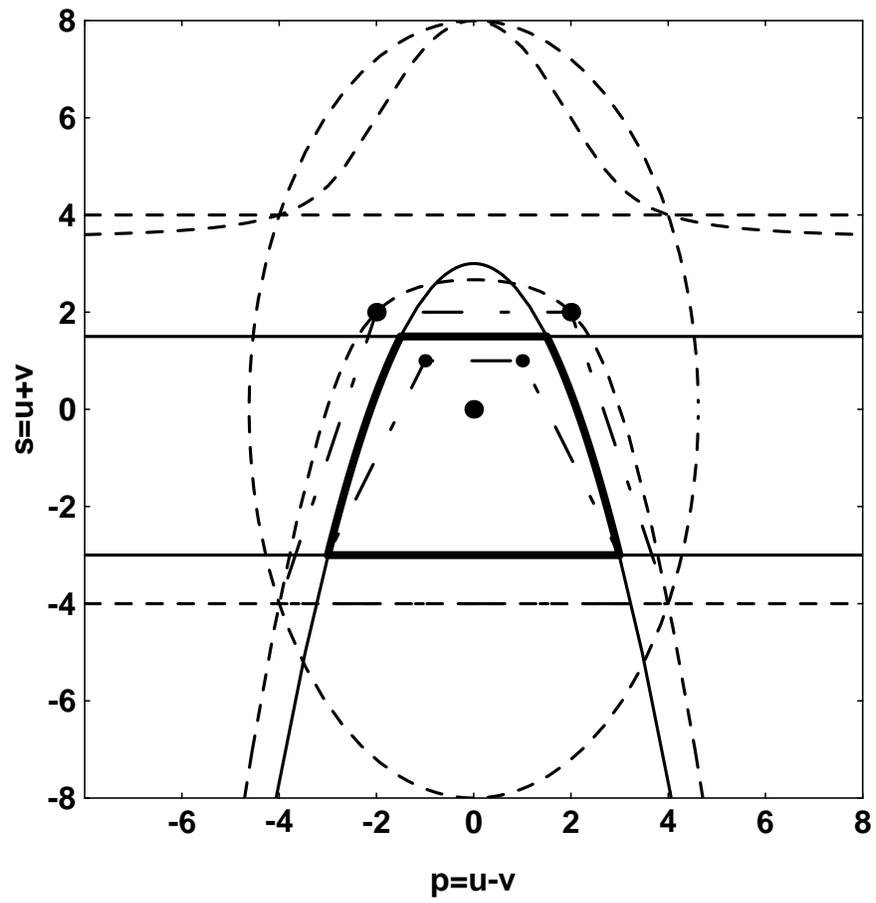
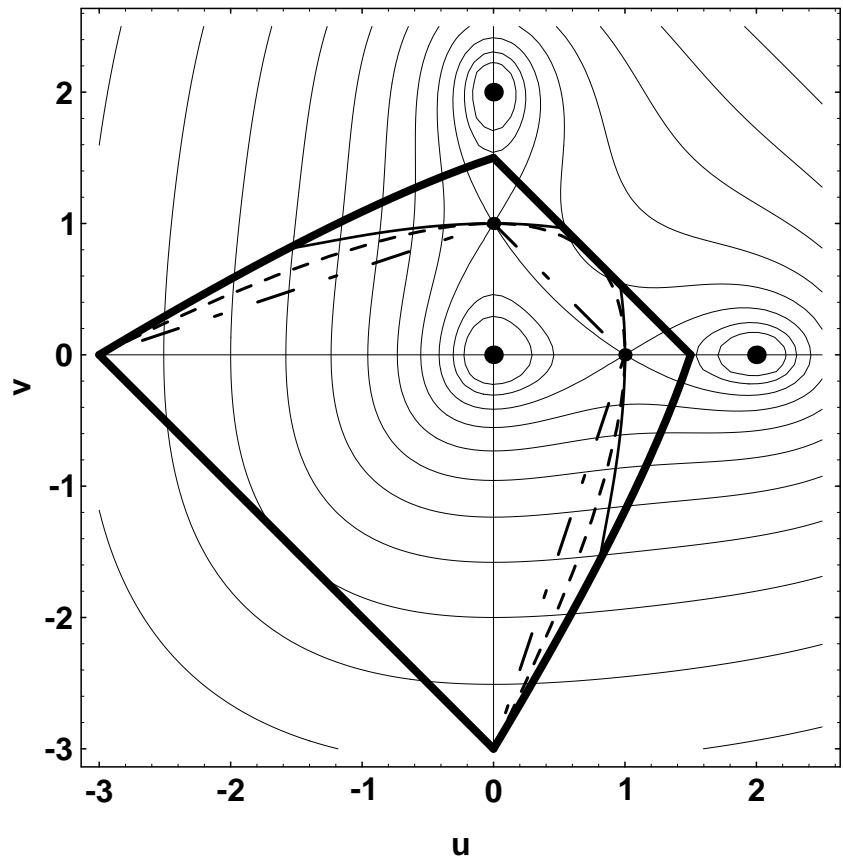
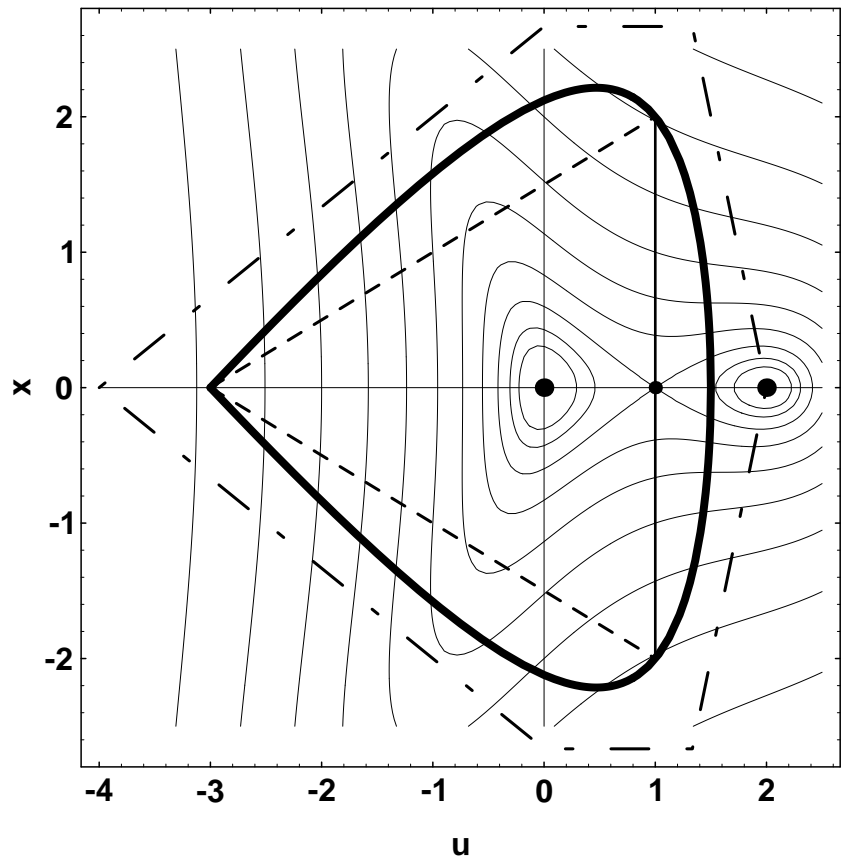
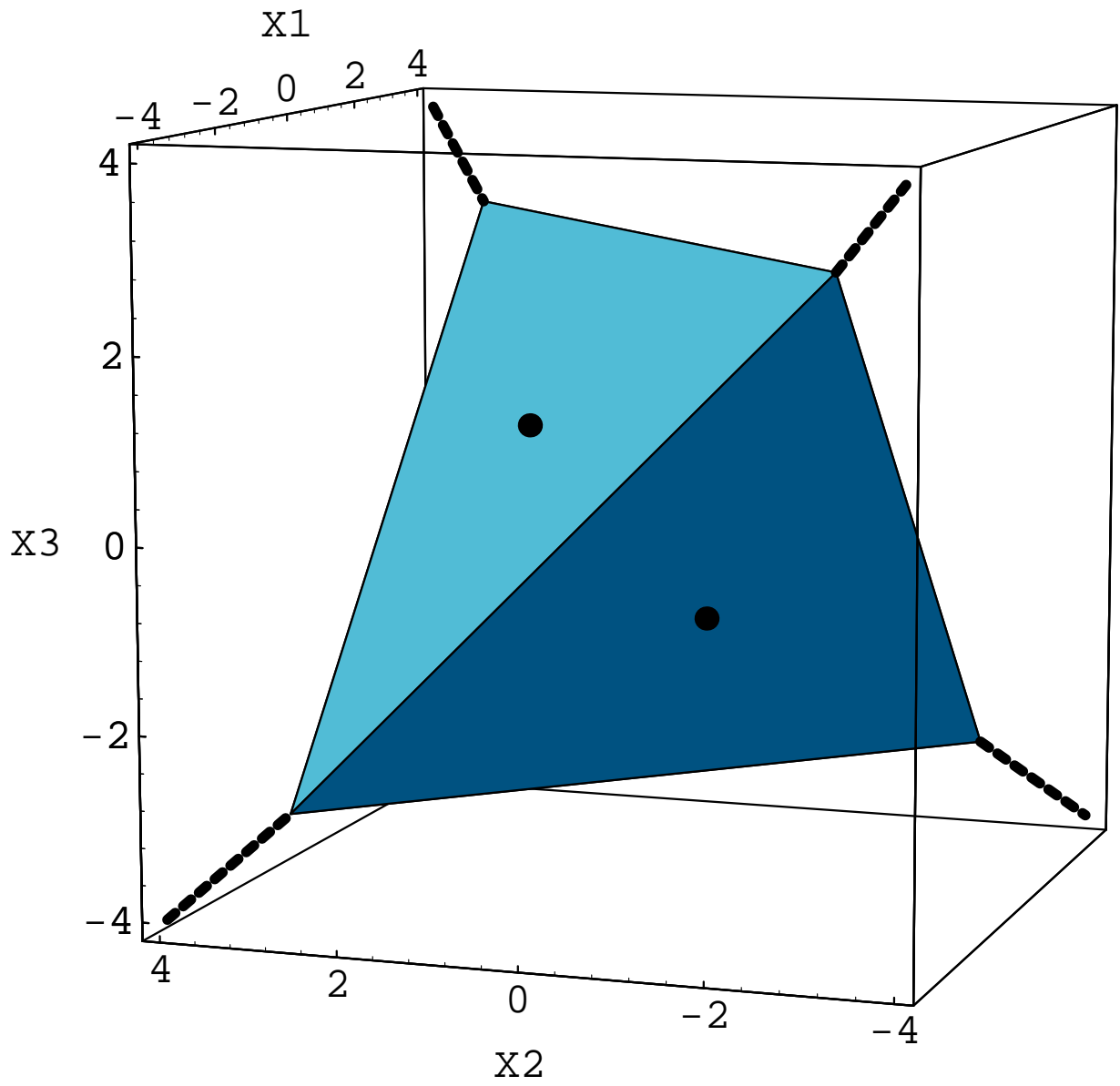
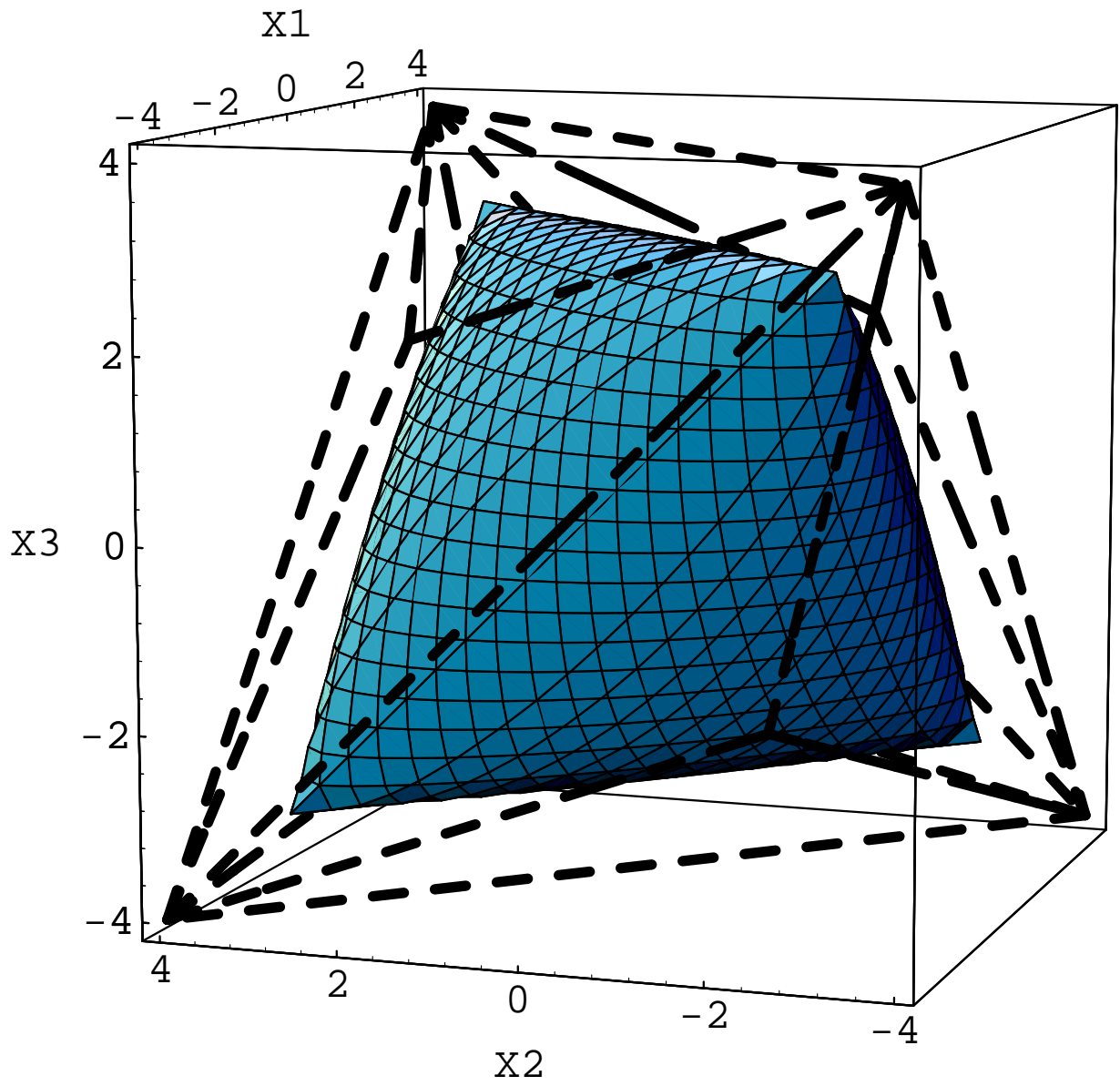
\end{document}